\documentclass[12pt,preprint]{aastex}
\usepackage{natbib}

\shorttitle{The Black Hole in PG\,2130+099}
\shortauthors{Grier et al.}

\begin{document}

\title{The Mass of the Black Hole in the Quasar PG\,2130+099}

\author{C.~J.~Grier\altaffilmark{1},
B.~M.~Peterson\altaffilmark{1,2},
M.~C.~Bentz\altaffilmark{3},
K.~D.~Denney\altaffilmark{1},
J.~D.~Eastman\altaffilmark{1},
M.~Dietrich\altaffilmark{1},
R.~W.~Pogge\altaffilmark{1,2},
J.~L.~Prieto\altaffilmark{1},
D.~L.~DePoy\altaffilmark{1,2},
R.~J.~Assef\altaffilmark{1},
D.~W.~Atlee\altaffilmark{1},
J.~Bird\altaffilmark{1},
M.~E.~Eyler\altaffilmark{4},
M.~S.~Peeples\altaffilmark{1},
R.~Siverd\altaffilmark{1},
L.~C.~Watson\altaffilmark{1}, 
J.~C.~Yee\altaffilmark{1}
}

\altaffiltext{1}{Department of Astronomy, The Ohio State University, 140 W 18th Ave,
Columbus, OH 43210}
\altaffiltext{2}{Center for Cosmology \& AstroParticle Physics, The Ohio State University, 
191 West Woodruff Ave, Columbus, OH 4321}
\altaffiltext{3}{Department of Physics and Astronomy, University of California at Irvine, 
4129 Frederick Reines Hall, Irvine, CA 92697}
\altaffiltext{4}{Department of Physics, U.S. Naval Academy, 572C Holloway Road, Annapolis, MD 21401}

\begin{abstract}
We present the results of a recent reverberation-mapping campaign undertaken to improve measurements
of the radius of the broad line region and the central black hole mass of the quasar PG\,2130+099. 
Cross correlation of the 5100\,\AA\ continuum and H$\beta$ emission-line light curves yields a 
time lag of 22.9$^{+4.4}_{-4.3}$ days, corresponding to a central black hole mass
$M_{\rm BH} = $ ($3.8 \pm 1.5 $)$ \times 10^{7} M_{\odot}$. This value supports the notion that previous measurements 
yielded an incorrect lag. We re-analyzed previous datasets to investigate the possible sources of the discrepancy and
conclude that previous measurement errors were apparently caused by a combination of undersampling of the light curves and
long-term secular changes in the H$\beta$ emission-line equivalent width. 
With our new measurements, PG\,2130+099 is no longer an outlier in either the $R_{\rm BLR}$--$L$ or 
the $M_{\rm BH}$--$\sigma_*$ relationships.  
\end{abstract}

\section{INTRODUCTION}
Reverberation mapping uses observations of continuum and emission-line variability to probe the 
structure of the broad line region (BLR) in active galactic nuclei (Blandford \& McKee 1982; Peterson 1993).
It has been extensively used to estimate the physical size of the BLR and the mass of central black holes 
in active galactic nuclei (AGN).  
The observed continuum variability precedes the observed emission-line variability by a time related to the
light travel time across the BLR; by obtaining an estimate of the time delay, or ``lag'' $\tau$, between 
the change in continuum flux and the change in emission-line flux, one can estimate the size of the BLR. 
While this is an extremely effective method, high-quality datasets are difficult to obtain, 
as they require well-spaced observations over long timescales. To date, over three dozen AGN
have estimated black hole masses obtained using reverberation methods. 

In addition to being important physical parameters, the BLR radius and central black hole mass ($M_{\rm BH}$) measurements are 
crucial in calibrating relationships between different properties of AGN. A useful 
relationship that has emerged is the correlation between the radius of the BLR ($R_{\rm BLR}$) and the optical 
luminosity of the AGN (e.g. Kaspi et al.\ 2000, 2005; Bentz et al.\ 2006). Another key relationship
is that between $M_{\rm BH}$ and bulge stellar velocity dispersion ($\sigma_*$) which is seen in both quiescent 
(Ferrarese \& Merritt 2000; Gebhardt et al.\ 2000a; Tremaine et al.\ 2002) and active galaxies (Gebhardt et al.\ 2000b; 
Ferraresse et al.\ 2001; Nelson et al.\ 2004; Onken et al.\ 2004; Dasyra et al.\ 2007). These two 
relationships are critical because they allow us to estimate the masses of black holes in AGN from a 
single spectrum (see McGill et al.\ 2008 for a recent summary) --- we estimate $R_{\rm BLR}$ from the AGN luminosity, 
the velocity dispersion of the 
BLR is determined by the emission-line width, and the quiescient $M_{\rm BH}$--$\sigma_{*}$ relationship provides the 
calibration of the reverberation-based mass scale. Obtaining high-quality single-epoch spectra is much less observationally 
demanding than obtaining high-quality reverberation measurements; once relations such as these are properly calibrated, we can 
measure $M_{\rm  BH}$ in many more objects than would otherwise be possible. An extensive set of objects with
reliable black hole masses allows us to explore the connection between black holes and AGN evolution over cosmologically
interesting timescales. 

PG\,2130+099 has been a source of curiosity because it is an outlier in both the $M_{\rm BH}$--$\sigma_{*}$ and $R_{\rm BLR}$--$L$ 
relations. Previous measurements obtained by Kaspi et al.\ (2000) and reanalyzed by Peterson et al.\ (2004) found H$\beta$ lags 
of about 180 days. 
These measurements yield a black hole mass $M_{\rm BH}$ upwards of $10^{8}$ $M_{\odot}$.
At luminosity $\lambda L_{\lambda}$(5100\,\AA)=($2.24\pm0.27$)$\times10^{44}$ erg s$^{-1}$, this places PG\,2130+099 well
above the $R_{\rm BLR}$--$L$ relationship (Bentz et al.\ 2006). Dasyra et al.\ (2007) also note that PG\,2130+099 
falls above the $M_{\rm BH}$--$\sigma_{*}$ relationship. Together these suggest that this discrepancy could be caused by 
measurement errors in the BLR radius. Our suspicions are also fueled by two other factors. First, the optical spectrum 
of PG\,2130+099 is similar to that of narrow-line Seyfert 1 (NLS1) galaxies, which are widely supposed to be AGN with high 
accretion rates relative to the Eddington rate (see Komossa 2008 for a recent review). However, the accretion rate derived 
from this mass and 
luminosity (under common assumptions for all reverberation-mapped AGN as described by Collin et al.\ 2006) is 
quite low compared to NLS1s, again suggesting that $M_{\rm BH}$ and therefore $\tau$ are overestimated. 
Second, as we discuss further below, a lag of approximately half a year on an equatorial source means that 
fine-scale structure in the two light curves will not match up in detail, and cross-correlation becomes very 
sensitive to long-term secular variations that may or may not be an actual reverberation signal. 

For these reasons, we decided to undertake a new reverberation campaign to remeasure the H$\beta$ lag for PG\,2130+099. 
In this paper, we present a new lag determination from 
this campaign. We also present a re-analysis of the earlier dataset that suggests that the true lag is consistent
with our new lag value and investigate possible sources of error in the previous analysis. 
 
\section{OBSERVATIONS AND DATA ANALYSIS}
 
\subsection{Observations}
The data were obtained as a part of queue-scheduled program during the SDSS-II Supernova Survey follow-up campaign 
(Frieman et al.\ 2008). We obtained spectra of PG\,2130+099 with the Boller and 
Chivens CCD Spectrograph on the MDM 2.4m telescope for 21 different nights from 2007 September through 2007
December. We used a 150 grooves/mm grating which yields a dispersion of 3.29 \AA\ pixel$^{-1}$ and 
covers the spectral range $4000$--$7500$ \AA. The slit width was set to $3''\!.0$ projected on the 
sky and the spectral resolution was 15.2\,\AA. We used an extraction aperture that corresponds to $3''\!.0 \times 7''\!.0$.
\subsection{Light Curves}
\begin{figure}
\begin{center}
\epsscale{0.8}
\plotone{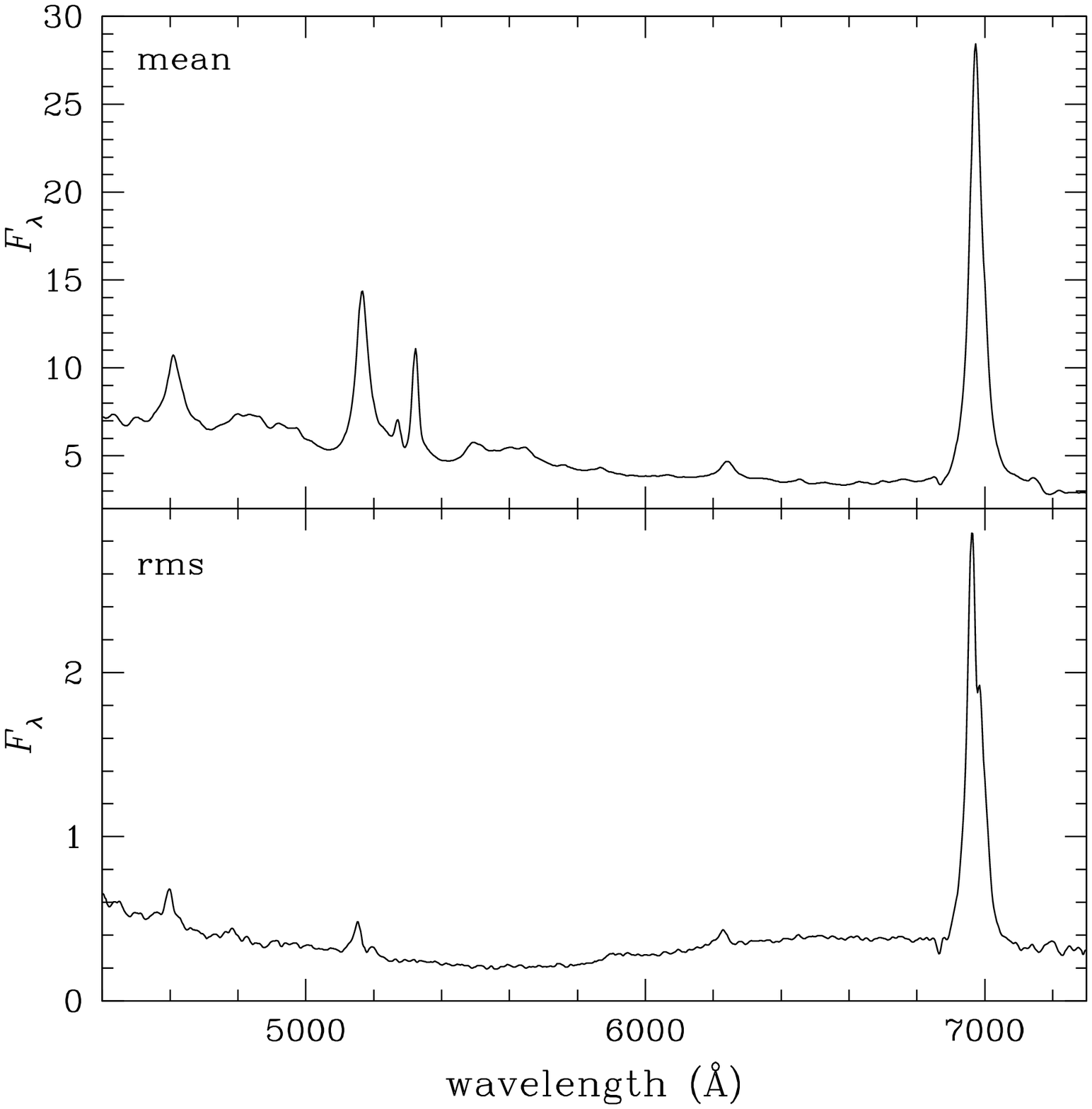}
\caption{The flux-calibrated mean and rms spectra of PG\,2130+099 in the observed frame ($z$ = 0.06298). 
The flux density is in units of 10$^{-15}$ erg s$^{-1}$ cm$^{-2}$ \AA$^{-1}$.}
\label{fig:f1}
\end{center}
\end{figure}

The reduced spectra were flux-calibrated using the flux of the [O\,{\sc iii}]\,$\lambda5007$ emission line in a reference 
spectrum created using a selection of 9 nights with the best observing conditions. Using a $\chi^2$ goodness-of-fit 
estimator method (van Groningen \& Wanders 1992), each individual spectrum was scaled 
to the reference spectrum. For three epochs, no reasonable fit could be obtained, hence we scaled these spectra 
by hand to match the [O\,{\sc iii}]\,$\lambda5007$ flux of the reference spectrum, which was 1.36$\times$10$^{-13}$ 
erg s$^{-1}$ cm$^{-2}$ in the observed frame ($z$~=~0.06298). We then removed the narrow-line 
components of the H$\beta$ and [O\,{\sc iii}]\,$\lambda\lambda\,4959$, $5007$ lines using relative line strengths given by 
Peterson et al. (2004). Figure \ref{fig:f1} shows the mean and rms spectra of PG\,2130+099, created from the entire set 
of 21 flux-calibrated observations. 

\begin{figure}
\begin{center}
\epsscale{0.9}
\plotone{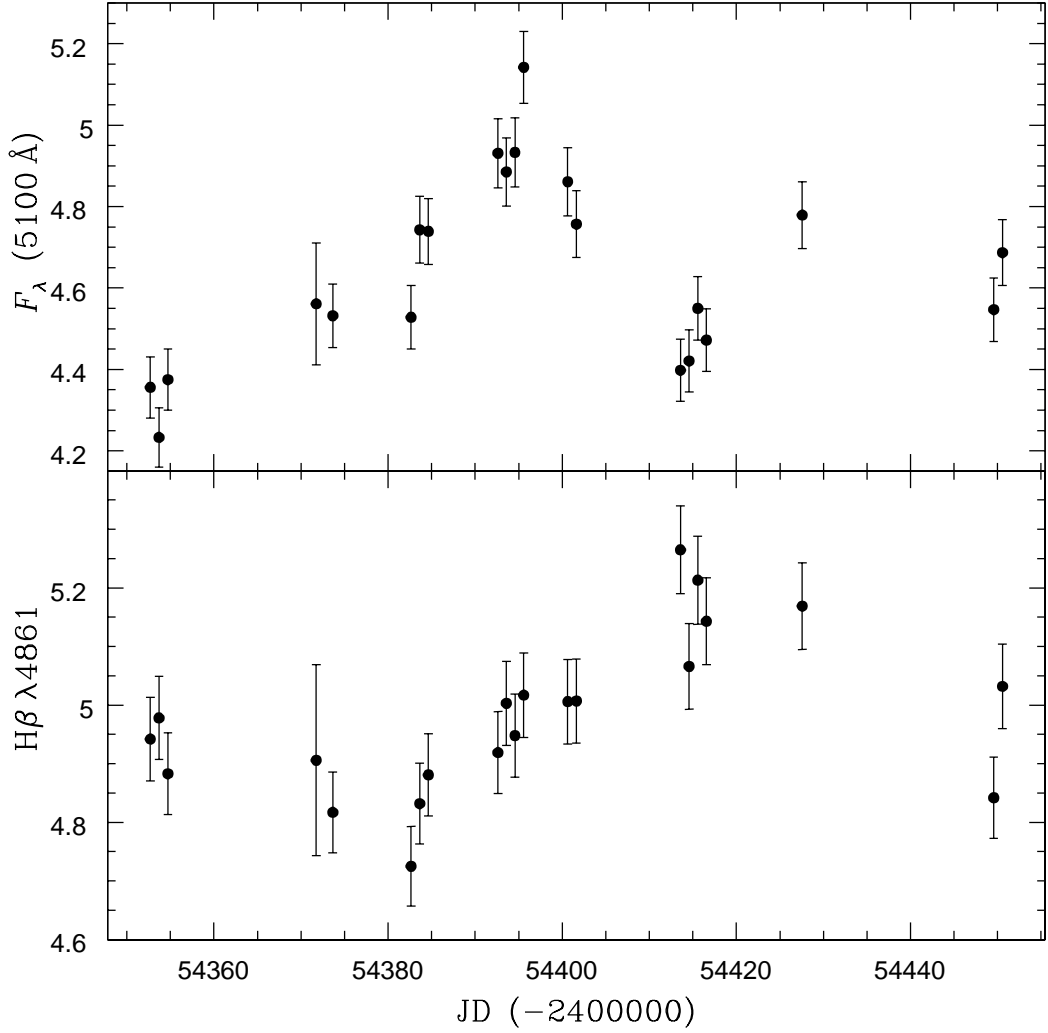}
\caption{Continuum (upper panel) and H$\beta$ (lower panel) light curves that were used in the time series analysis. Continuum flux 
densities are in units of 10$^{-15}$ erg s$^{-1}$ cm$^{-2}$ \AA$^{-1}$. Emission-line flux densities are 
in units of 10$^{-13}$ erg s$^{-1}$ cm$^{-2}$. All fluxes are in the observed frame.}
\label{fig:f2}
\end{center}
\end{figure}

To measure the continuum and line fluxes, we fit a line to the continuum between the regions $5050$--$5070$\,\AA\ 
and $5405$--$5435$\,\AA\ in 
the observed frame. We measured the H$\beta$ flux by integrating the flux above the continuum from $5070$--$5285$\,\AA,
and the optical continuum flux was taken to be the average flux in the range $5405$--$5435$\,\AA. Errors were estimated 
based on differences between observations that were close together in time. The observations from 2007 
September 27 (JD\,2454371) had an extremely low signal-to-noise ratio $(S/N)$ compared to the rest of the population; following 
Peterson et al.\ (1998a), we assigned fractional errors of 1/$(S/N)$ for this night in both the continuum 
and line flux. The resulting continuum and H$\beta$ light curves are given in Table \ref{Table:tbl1} and 
are shown in Figure \ref{fig:f2}. The properties of the final light curves used in our time series analysis 
are given in Table \ref{Table:tbl2}. Column (1) gives the spectral feature and Column (2) gives the number of points in the individual
light curves. Columns (3) and (4) give the average and median time spacing between observations, respectively. Column (5) 
gives the mean flux of the feature in the observed frame, and Column (6) gives the mean fractional error that is 
computed based on observations that are closely spaced. Column (7) gives the excess variance, defined by 
\begin{equation}
F_{\rm var} = \frac{\sqrt{\sigma^{2}-\delta{^2}}}{\langle f\rangle}
\end{equation}
where $\sigma^{2}$ is the flux variance of the observations, $\delta^{2}$ is the mean square uncertainty, and $\langle f \rangle$
is the mean observed flux. Column (8) is the ratio of the maximum to minimum flux in each light curve.

We note in passing that we attempted to determine light curves for the other prominent emission lines in our spectra, 
H$\alpha$ and H$\gamma$. Unfortunately, the fidelity of our flux calibration decreases away from the 
[O\,{\sc iii}]\,$\lambda5007$ line (which we believe accounts for the increasing strength of the rms spectrum longward of 
$\sim 5800$\,\AA\ in Figure \ref{fig:f1}) and, combined with the low amplitude of variability, renders the other light curves 
unreliable. Nevertheless, time series analysis yields results that are at least consistent with the H$\beta$ results, though with 
much larger uncertainties.

\section {TIME SERIES ANALYSIS}
\subsection {Time delay measurements}
\begin{figure}
\begin{center}
\epsscale{0.75}
\plotone{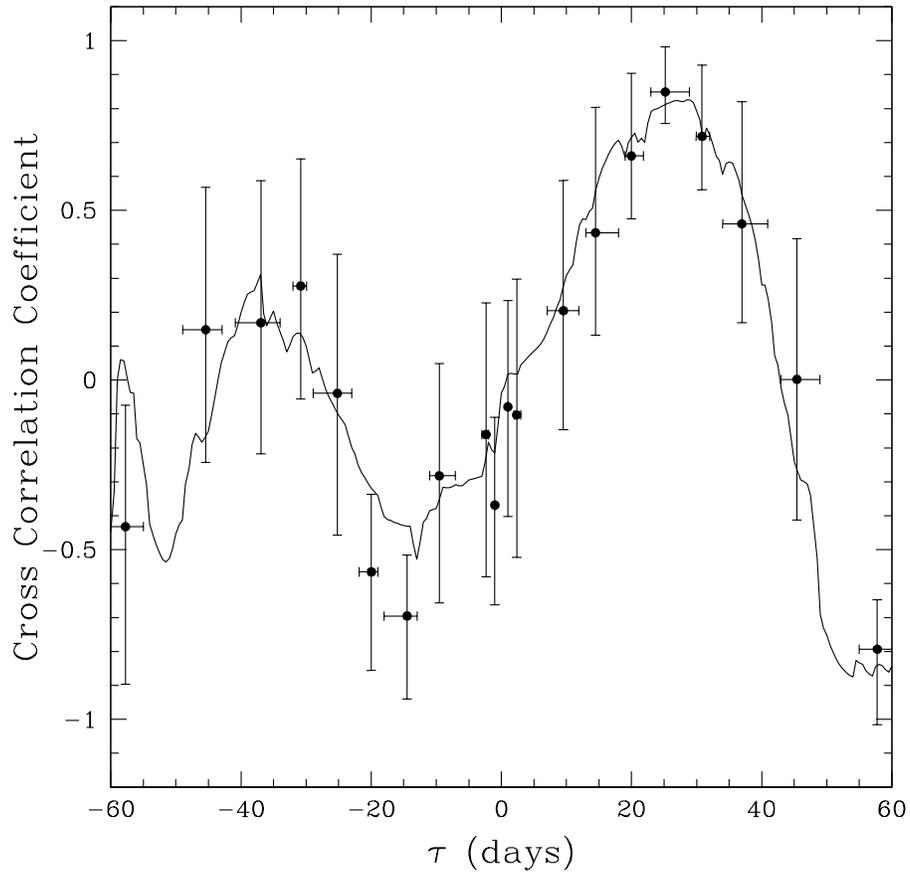}
\caption{The CCF (solid line) and ZDCF (filled circles) from the time series analysis of 
our recent observations of PG\,2130+099.}
\label{fig:f3}
\end{center}
\end{figure}

To measure the lag between the optical continuum and H$\beta$ emission-line variations, we cross-correlated the 
H$\beta$ light curve with the continuum light curve using the interpolation method originally described by 
Gaskell \& Sparke (1986) and Gaskell \& Peterson (1987) and subsequently modified by White \& Peterson (1994) and 
Peterson et al.\ (1998b, 2004). The method works as follows: because the data 
are not evenly spaced in time, we interpolate between points to obtain an evenly-sampled light curve. 
The linear correlation coefficient $r$ is then calculated using pairs of points, one from each light curve, 
separated by a given time lag. When $r$ is calculated for a range of time lag values, the cross correlation function 
(CCF) is obtained, which consists of the value of $r$ for each time lag value. We interpolated 
between the points on a 0.5 day timescale to obtain the CCF for our dataset, which is shown in Figure \ref{fig:f3}.

To determine the most probable delay and its uncertainty, we employed the flux randomization/random subset sampling 
Monte Carlo method described by Peterson et al.\ (1998b). The ``random subset sampling'' method 
takes a light curve of $N$ points and samples it $N$ times without regard to whether or not any given point has been 
selected already. The flux uncertainty of a data point selected $n$ times is correspondingly 
reduced by a factor $n^{-1/2}$ (Welsh 1999; Peterson et al. 2004). The flux value of each point is then altered by 
a random Gaussian deviate based on the uncertainty assigned to the point; this is known as ``flux randomization''. A 
CCF was calculated for each 
altered light curve using interpolation as before. Using 2000 iterations of this process, we obtain a distribution of 
$\tau_{\rm peak}$, measured from the CCF of each Monte Carlo realization and defined by the location of the peak value 
of $r$. We also calculate and obtain a distribution
of the lag values $\tau_{\rm cent}$ that represents the centroid of each CCF using the points surrounding the peak, based on 
all lag values with $r\geq0.8r_{\rm max}$. We then adopt the mean values of both the centroid 
and peak $\langle \tau_{\rm cent}\rangle$ and $\langle \tau_{\rm peak}\rangle$ for our analysis. We estimate the uncertainties
in $\tau_{\rm cent}$ and $\tau_{\rm peak}$ such that 15.87\% of the realizations fall below the mean minus the lower 
error and 15.87\% of the 
realizations fall above the mean plus the upper error. Our final lag values and errors are $\tau_{\rm cent}=22.9^{+4.4}_{-4.3}$ 
days and $\tau_{\rm peak}=22.2^{+5.6}_{-5.2}$ days in the rest frame of PG\,2130+099. It is important to note that the CCF
is heavily dependent on the timespan of the light curve relative to the actual delay, and if the data timespan is short 
with respect to the lag, the CCF will be biased towards short lag measurements (Welsh 1999). We note that because 
our data span just over three months, our CCF is not capable of producing a lag measurement as large as previous measurements and
is subject to this bias. However, if the actual lag is shorter as the evidence presented in this work suggests, our 
data span a sufficiently long period to identify the true delay. 

As a check on our time series analysis, we also employed an alternative method known as the $z$-transformed 
discrete correlation function (ZDCF), as described by Alexander (1997). This method is a modification of the discrete 
correlation function (DCF) described by Edelson \& Krolik (1988). The DCF is obtained by correlating data points 
from the continuum with points from the H$\beta$ light curve by binning the data in time rather than interpolating 
between points. The ZDCF similarly uses bins rather than interpolation, but  bins the data by equal population rather 
than by equal spacing in time and applies Fisher's $z$-transform to the cross correlation coefficients. Discrete 
correlation functions have the advantage that they only use real data points, and are therefore less likely to 
find spurious correlations in data where there are large time gaps. However, datasets with few points tend to 
yield lag values with large uncertainties; hence this method is primarily useful as a check on the interpolation 
method when the data are undersampled. Figure \ref{fig:f3} shows the computed ZDCF for our PG\,2130+099 dataset. It 
should be noted that the $z$-transform requires a minimum of 11 points per bin to be statistically significant (Alexander
1997); 
because our light curves contained only 21 points, we were unable to obtain a well-sampled ZDCF with a minimum of 11 points 
per bin. We used a minimum of 8 points per bin and do not obtain an independent lag measurement, but the ZDCF 
is very consistent with our calculated CCF, indicating that the interpolation results are credible.

\subsection {Line width measurement and mass calculations}

We use the second moment of the line profile, $\sigma_{\rm line}$ (Fromerth \& Melia 2000; Peterson et al.\ 2004) 
to characterize the width of the H$\beta$ line. To determine the best value of $\sigma_{\rm line}$ and 
its uncertainties, we use Monte Carlo simulations similar to those used in determining the time lag 
(see Peterson et al. 2004). We measure the line widths from both the mean and rms spectra created in the 
simulations from $N$ randomly chosen spectra and correct them for the spectrograph resolution. 
We obtain a distribution of each line width from multiple realizations, 
from which we take the mean value for our measure of $\sigma_{\rm line}$ or FWHM and the standard deviation as
the measurement uncertainty. The measured line widths and uncertainties are given in Table \ref{Table:tbl3}. 

Assuming that the motion of the H$\beta$-emitting gas is dominated by gravity, the relation 
between $M_{\rm BH}$, line width, and time delay is
\begin{equation}
M_{\rm BH} = \frac{f c \tau \Delta V^2}{G},
\end{equation}
where $\tau$ is the emission-line time delay, $\Delta V$ is the emission-line width, and $f$ 
is a dimensionless factor that is characterized by the geometry and kinematics of the BLR. Onken 
et al.\ (2004) calculated an average value of $\langle f \rangle$ by assuming that AGN follow the 
same $M_{\rm BH}$--$\sigma_*$ relationship as quiescent galaxies. By normalizing the reverberation black 
hole masses to the relation for quiescent galaxies, they obtain $\langle f\rangle =5.5$ when the line width
is characterized by the line dispersion $\sigma_{\rm line}$ of the rms spectrum.\footnote{Differences 
between the use of $\sigma_{\rm line}$ and the FWHM in calculating $M_{\rm BH}$ are discussed 
by Collin et al.\ (2006).} Using our values of $\tau_{\rm cent}$ 
and $\sigma_{\rm line}$, we compute $M_{\rm BH} = $ ($3.8 \pm 1.5 $)$ \times 10^{7} M_{\odot}$.

We also compute the average 5100\,\AA\ luminosity of our sample to see where our new $R_{\rm BLR}$ value 
places PG\,2130+099 on the $R_{\rm BLR}$--$L$ relationship. Following Bentz et al.\ (2006), we correct our 
luminosity for host-galaxy contamination. Assuming $H_{0}=70$ km s$^{-1}$ 
Mpc$^{-1}$, $\Omega_{\rm m}=0.30$ and $\Omega_{\rm \Lambda}=0.70$, we 
calculate $\log \lambda L_{\lambda}$(5100\,\AA)= 44.40$\pm0.02$
for our recent measurements of PG\,2130+099. The relevant measurements 
and computed quantities are given in Table \ref{Table:tbl4}. 

\section{DISCUSSION}
\begin{figure}
\begin{center}
\epsscale{1.0}
\plotone{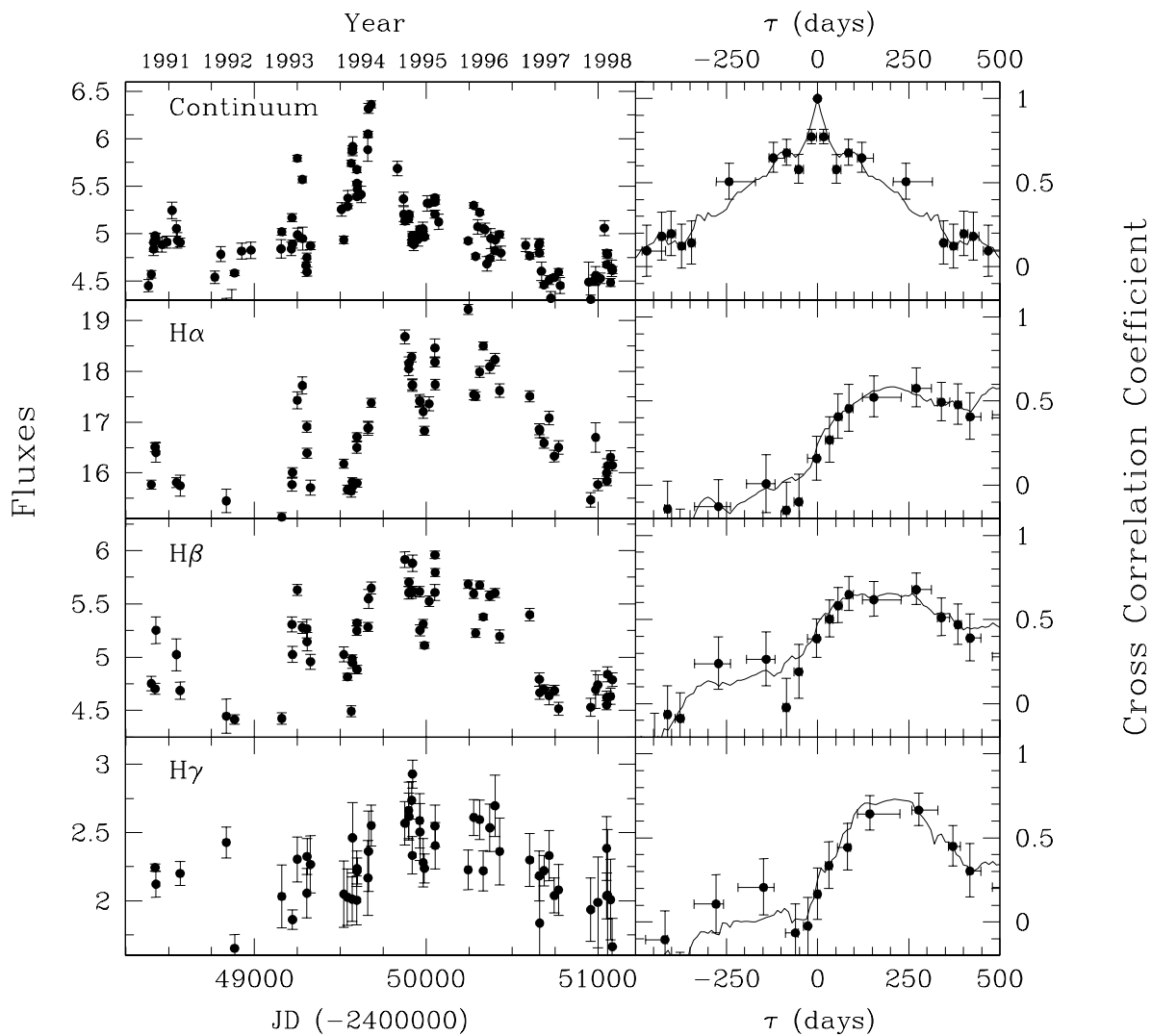}
\caption{The left column shows the light curves of PG\,2130+099 from Kaspi et al.\ (2000). The 5100\ \AA \ continuum light curve is shown 
in the top left panel and the H$\alpha$, H$\beta$, and H$\gamma$ light curves are displayed below. The right column 
shows the CCF and corresponding ZDCF for each spectral feature, with the top right panel showing the auto-correlation 
function (ACF) of the continuum. The continuum flux density is given in units of 10$^{-15}$ erg s$^{-1}$ 
cm$^{-2}$~\AA$^{-1}$ and the H$\beta$ flux density is in units of 10$^{-13}$ erg s$^{-1}$ cm$^{-2}$. }
\label{fig:f4}
\end{center}
\end{figure}

Analysis of previous datasets of PG\,2130+099 yields lags greater than 150 days and $M_{\rm BH}$  
values above $10^{8}$ $M_{\odot}$, in both cases approximately an order of magnitude larger than our new value. 
To investigate the source of these discrepancies, we closely scrutinized the previous dataset, 
which consists of data obtained at Steward Observatory and Wise Observatory (Kaspi et al.\ 2000).
The 5100\,\AA\ continuum and emission-line light curves, along with their respective CCF and ZDCFs, are shown in Figure \ref{fig:f4}. 
We first ran the full light curve through the time series analysis as described in section 2.3 to 
confirm the previous results, and successfully reproduced a lag measurement of $\sim168$ days, in agreement 
with Kaspi et al.\ (2000) and Peterson et al.\ (2004). However, visual inspection of the light curves 
suggests that these values were in error; we were able to identify several features that are 
present in the continuum, H$\beta$ and H$\alpha$ light curves that are quite visibly lagging on timescales 
shorter than 50 days, as we show below. 

\begin{figure}
\begin{center}
\epsscale{1.0}
\plotone{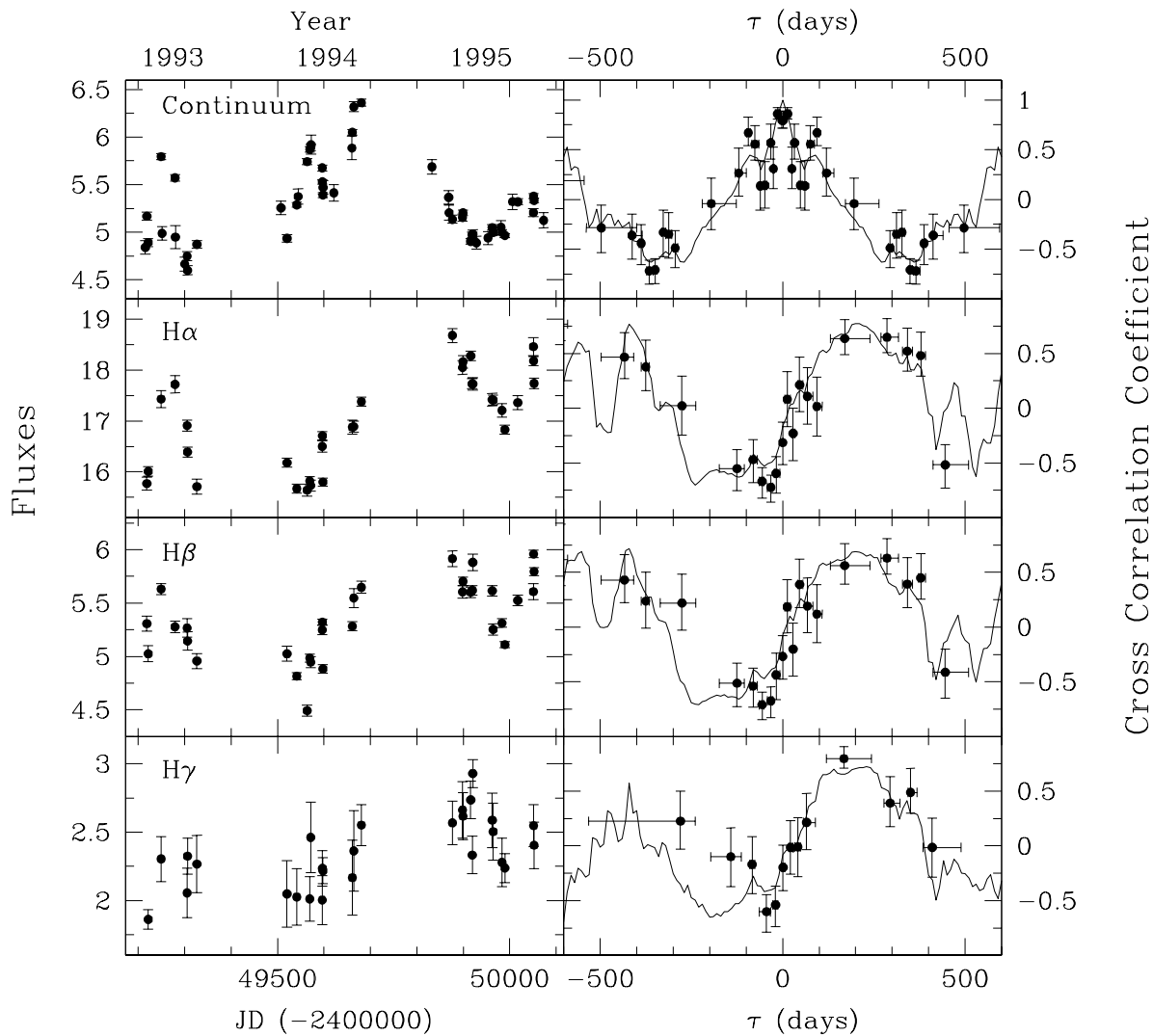}
\caption{Light curves and cross correlation functions for different spectral features of the Kaspi et al.\ (2000) data.
The left column shows a small section of the light curve for each spectral feature. The right column panels show the CCF (solid line)
and ZDCF (filled circles) for each feature. The top right panel shows the ACF of the continuum, and the other three panels 
below show the correlation functions resulting from correlation of each respective emission line with the continuum 
light curve. The fluxes are given in the same units as in Figure \ref{fig:f4}.}
\label{fig:f5}
\end{center}
\end{figure}

It is clear that the Kaspi et al.\ results are dominated by the 
data spanning the years 1993-1995, as this period contains the most significant variability as well 
as time sampling that is sufficient to resolve features in the light curves. We show the light curves for
all spectral features measured by Kaspi et al.\ during this time period in Figure \ref{fig:f5}. Over 
this period, we can match behavior 
in the continuum with the behavior of the emission lines, most noticeably in the data from 1995. The 
most prominent features in the light curves are the maximum in the continuum at the end of the 1994 series
and the maxima in the lines at the beginning of the 1995 series, which, based on the continuum variations, one 
would expect to be lower relative to the fluxes in 1994. However, inspection of the relative flux levels of the lines
and continuum in the 1994 and 1995 data reveals that the equivalent width of the emission lines 
changes during this timespan, which results in a rise in emission-line flux apparently unrelated to 
reverberation. The maximum in the continuum and those 
in the emission lines do not correspond to the same features--- the maxima in the emission line light curves in 1995 correspond 
to the similar feature in the continuum at the beginning of 1995. Because there are no observations during 
the six months or so between these features, the CCF and ZDCF lock onto the two unrelated maxima that are separated by 196
days and yield lags 
of $\sim$200 days over this three-year span. The emission lines likely continued to increase in flux during the time period 
for which there were no observations. 

Inspection of the light curve segments in Figure \ref{fig:f5} reveals similar structures in each of them within a given year. 
To quantify the time delays between the continuum and lines on short timescales, we cross-correlated each individual year, 
this time using only flux randomization in the Monte Carlo realizations, as there were too few points 
in each year to use random subset sampling, so the uncertainties are underestimated. The resulting lags are 
given in Table \ref{Table:lags}. Again we must 
consider that the CCF cannot produce a lag that is greater than the duration of observations, so the 
individual CCFs for these three years are limited to short delays. However, the presence of the 1995 feature in 
both the continuum and emission-line light curves is unmistakable and it is extremely unlikely that it lags 
by a value exceeding the sensitivity limit of the CCF. From this we surmise that the discrepancies in measured lag values 
are mostly a result of large time gaps in the data and/or underestimation of the error bars in the Kaspi et al.\ data. 

\begin{figure}
\begin{center}
\epsscale{1.0}
\plotone{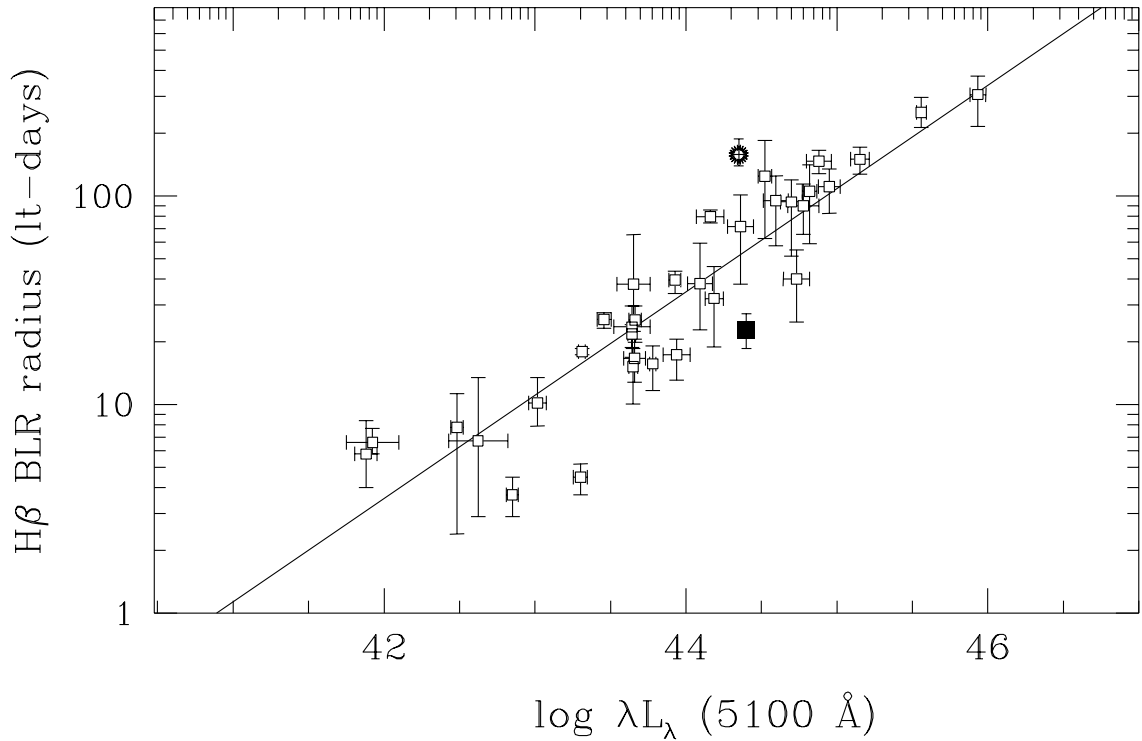}
\caption{The position of PG\,2130+099 on the most recent $R_{\rm BLR}$--$L$ relationship from Bentz 
et al.\ (2008), denoted by the solid line. The starred circle represents the position of 
PG\,2130+099 with a lag of 158.1 days, as given by Peterson et al.\ (2004). The filled square represents 
the measurement of 22.9 days from our recent dataset. The open squares represent other 
objects from Bentz et al. }
\label{fig:f6}
\end{center}
\end{figure}

\begin{figure}
\begin{center}
\epsscale{1.0}
\plotone{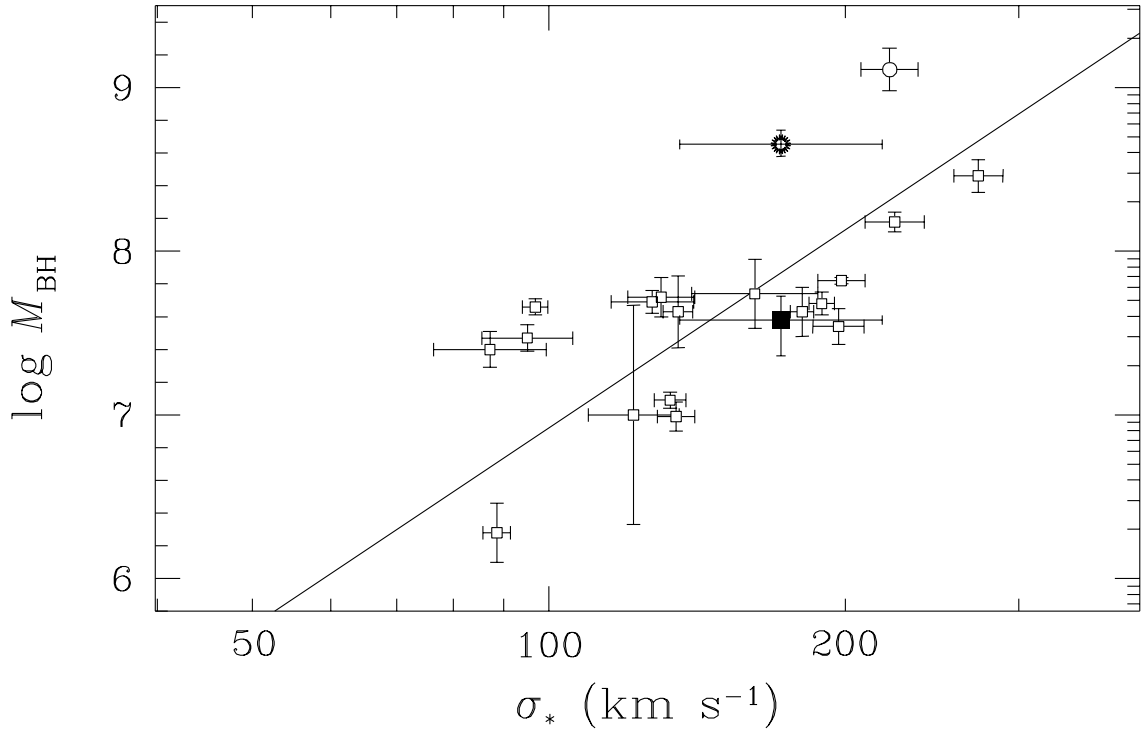}
\caption{The position of PG\,2130+099 on the $M_{\rm BH}$--$\sigma_{*}$ relationship (from Tremaine et al.\ 2002), denoted
by the solid line. The 
open squares show data points based on Onken et al.\ (2004) and Nelson et al.\ (2004) and the open circle is PG\,1426+015 from 
Watson et al.\ (2008). The starred circle and filled square represent the position of PG\,2130+099 using the 
same data sets as in Figure \ref{fig:f6}.}
\label{fig:f7}
\end{center}
\end{figure}

Welsh (1999), and even earlier, P\'{e}rez, Robinson, \& de la Fuente (1992), pointed out that emission-line lags can 
be severely underestimated with light curves that are too short in duration, particularly if the BLR is extended: 
certainly our 98-day campaign is insensitive to lags as long as $\sim180$\,days. Is it possible that the original lag 
determination of Kaspi et al.\ (2000) and Peterson et al.\ (2004) is correct (or more nearly so) and we have been fooled 
by reliance on light curves that are too short? We think not, based on (1) the reasonable match between details in the 
continuum and emission-line light curves for four different observing seasons (1993, 1994, 1995, and 2007), (2) the 
improved agreement with the $R_{\rm BLR}$-$L$ relationship with the smaller lag (demonstrated in Figure \ref{fig:f6}), and (3) the 
improved agreement with the $M_{\rm BH}$-$\sigma_*$ relationship with the smaller lag (Figure \ref{fig:f7}). It is also worth 
pointing out the difficulty of accurately measuring a $\sim 180$\,day lag, particularly in the case of equatorial sources 
which have a relatively short observing season. The short observing season, typically 6-7 months, means that there are 
very few emission-line observations that can be matched directly with continuum points: the observed emission-line 
fluxes represent a response to continuum variations that occurred when the AGN was too close to the Sun to observe.

Welsh (1999) has also pointed out the value in ``detrending'' the light curves--- removing long-term trends by fitting the 
light curves with a low-order function can reduce the bias toward underestimating lags. In this particular case, we find that 
detrending has almost no effect: in particular, the highest points in the continuum (in late 1994) and the highest points 
in the line (early in the 1995 observing season) remain so after detrending, and both the interpolation CCF and ZDCF 
weight these points heavily. It is also interesting to note that the peak in the ZDCF agrees with the peak of the 
interpolation CCF (Figure \ref{fig:f4}), which demonstrates that the $\sim180$\,day lag is {\em not} simply ascribable to 
interpolation across the gap between the 1994 and 1995 observing seasons.

The data from our 2007 campaign are not ideal: the amplitude of variability was low, the time sampling was 
adequate, but only barely, and the duration of the campaign was short enough that we lack sensitivity to lags of $\sim 
50$\,days or longer. But the preponderance of evidence at this point argues that our smaller lag measurement is more 
likely to be correct than the previous determination. Certainly better-sampled light curves of longer duration would yield a 
more definitive result.

\section {SUMMARY}
From a new reverberation campaign, we obtain a measurement of the time lag 
of the H$\beta$ line in PG\,2130+099 of $\tau_{\rm cent}$= 22.9$^{+4.4}_{-4.3}$ days 
and calculate a central black hole mass $M_{\rm BH} = $ ($3.8 \pm 1.5 $)$ \times 10^{7} M_{\odot}$.  
Previous measurements of $\tau_{\rm cent}$ overestimated the size of the BLR and therefore 
the mass of the black hole, most likely due to the undersampling in the light curves. 
A re-analysis of the previous data, using both visual and algorithmic methods, suggests that the true BLR size 
and $M_{\rm BH}$ are consistent with our new values. The recent reverberation measurements for PG\,2130+099 presented 
in this study remove the discrepancies previously found for this object based on the $M_{\rm BH}$--$\sigma_*$ 
and $R_{\rm BLR}$--$L$ relationships.

\acknowledgments

We are grateful for support of this program by the National Science Foundation through
grant AST-0604066 to The Ohio State University. We would also like to thank H. Netzer, S. Kaspi, 
and referee C. M. Gaskell for their helpful suggestions.

\clearpage

\clearpage

\clearpage

\clearpage

\begin{deluxetable}{lccc} 
\tablewidth{0pt} 
\tablecaption{Continuum and H$\beta$ Fluxes for PG\,2130+099} 
\tablehead{ 
\colhead{JD} & 
\colhead{$F_{\lambda}$ (5100\,\AA )} &
\colhead{H$\beta$\,$\lambda4861$} \\ 
\colhead{($-2400000$)} & 
\colhead{($10^{-15}$ erg s$^{-1}$ cm$^{-2}$ \AA$^{-1}$) } & 
\colhead{ ($10^{-13}$ erg s$^{-1}$ cm$^{-2}$)} & 
} 
\startdata
54352.70 &  4.36 $\pm $ 0.08&   4.94 $\pm $ 0.07\\
54353.70 &  4.23 $\pm $ 0.07&   4.98 $\pm $ 0.07\\
54354.72 &  4.38 $\pm $ 0.08&   4.88 $\pm $ 0.07\\
54371.76 &  4.56 $\pm $ 0.15&   4.91 $\pm $ 0.16\\
54373.65 &  4.53 $\pm $ 0.08&   4.82 $\pm $ 0.07\\
54382.65 &  4.53 $\pm $ 0.08&   4.73 $\pm $ 0.07\\
54383.63 &  4.74 $\pm $ 0.08&   4.83 $\pm $ 0.07\\
54384.64 &  4.74 $\pm $ 0.08&   4.88 $\pm $ 0.07\\
54392.62 &  4.93 $\pm $ 0.09&   4.92 $\pm $ 0.07\\
54393.59 &  4.89 $\pm $ 0.08&   5.00 $\pm $ 0.07\\
54394.57 &  4.93 $\pm $ 0.08&   4.95 $\pm $ 0.07\\
54395.59 &  5.14 $\pm $ 0.09&   5.02 $\pm $ 0.07\\
54400.63 &  4.86 $\pm $ 0.08&   5.01 $\pm $ 0.07\\
54401.63 &  4.76 $\pm $ 0.08&   5.01 $\pm $ 0.07\\
54413.57 &  4.40 $\pm $ 0.08&   5.27 $\pm $ 0.08\\
54414.58 &  4.42 $\pm $ 0.08&   5.07 $\pm $ 0.07\\
54415.57 &  4.55 $\pm $ 0.08&   5.21 $\pm $ 0.08\\
54416.56 &  4.47 $\pm $ 0.08&   5.14 $\pm $ 0.07\\
54427.57 &  4.78 $\pm $ 0.08&   5.17 $\pm $ 0.07\\
54449.56 &  4.55 $\pm $ 0.08&   4.84 $\pm $ 0.07\\
54450.57 &  4.69 $\pm $ 0.08&   5.03 $\pm $ 0.07
\enddata
\label{Table:tbl1}
\end{deluxetable}

\begin{deluxetable}{lccccccc}
\tablewidth{0pt}
\tablecaption{Light Curve Statistics}
\tablehead{
\colhead{ } &
\colhead{ } &
\multicolumn{2}{c}{Sampling} &
\colhead{ } &
\colhead{Mean} \\
\colhead{Time} &
\colhead{ } &
\multicolumn{2}{c}{Interval (days)} &
\colhead{Mean} &
\colhead{Fractional} \\
\colhead{Series } &
\colhead{$N$} &
\colhead{$\langle T \rangle$} &
\colhead{$T_{\rm median}$} &
\colhead{Flux\tablenotemark{1}} &
\colhead{Error} &
\colhead{$F_{\rm var}$} &
\colhead{$R_{\rm max}$} \\
\colhead{(1)} &
\colhead{(2)} &
\colhead{(3)} &
\colhead{(4)} &
\colhead{(5)} &
\colhead{(6)} &
\colhead{(7)} &
\colhead{(8)} 
} 
\startdata
5100\,\AA\ & 21 & 4.9 & 1.0 & $4.64\pm0.23$ & 0.018 & 0.047 & $1.22\pm 0.03$\\
H$\beta$ & 21 & 4.9 & 1.0 & $4.98\pm0.14$ & 0.015 & 0.023 & $1.14\pm 0.02$
\enddata
\tablenotetext{1}{Continuum and emission-line fluxes are given in the same units
as Table \ref{Table:tbl1}.}
\label{Table:tbl2}
\end{deluxetable} 

\begin{deluxetable}{lcc} 
\tablewidth{0pt} 
\tablecaption{Reverberation Results} 
\tablehead{ 
\colhead{Parameter} & 
\colhead{Value} \\ 
\colhead{(1)} & 
\colhead{(2) }  
} 
\startdata
$\tau_{\rm cent} $\tablenotemark{a}	      &  22.9 $^{+4.7}_{-4.6}$ days\\
$\tau_{\rm peak} $\tablenotemark{a}	      &  22.2 $^{+5.6}_{-5.2}$ days \\
$\sigma_{\rm line}$ (mean)                    &  1807 $\pm$ 4 km s$^{-1}$\\
FWHM (mean) 	                              &	 2807 $\pm$ 4 km s$^{-1}$\\
$\sigma_{\rm line}$ (rms)                     &	 1246 $\pm$ 222  km s$^{-1}$\\
FWHM (rms)	                              &	 2063 $\pm$ 720 km s$^{-1}$\\
$M_{\rm BH}$              	              &  ($3.8 \pm 1.5 $)$ \times 10^{7} M_{\odot}$
\enddata
\tablenotetext{a}{Values are given in the rest frame of the object.}
\label{Table:tbl3}
\end{deluxetable} 

\begin{deluxetable}{lcc} 
\tablewidth{0pt} 
\tablecaption{Host Galaxy Flux Removal Parameters} 
\tablehead{ 
\colhead{Parameter} & 
\colhead{Value} \\ 
\colhead{(1)} & 
\colhead{(2) }  
} 
\startdata
Position angle of slit      & 0.0$^{\circ}$\\
Aperture size               &$ 3''\!.0 \times 7''\!.0$\\
$F_{\rm galaxy}$ (5100\,\AA)\tablenotemark{a} & (0.405$\pm$0.037) $\times 10^{-15}$ erg s$^{-1}$ cm$^{-2}$ \AA$^{-1}$\\
$\log\lambda$L$_{\lambda}$(5100\,\AA)\tablenotemark{b}  & (44.40$\pm0.02$) 
\enddata
\tablenotetext{a}{Galaxy flux is in the observed frame, from Bentz et al.\ (2008).}
\tablenotetext{b}{5100\,\AA\ rest-frame luminosity in erg s$^{-1}$, with host galaxy subtracted and corrected for extinction, 
following Bentz et al.\ (2006; 2008).}
\label{Table:tbl4}
\end{deluxetable}


\begin{deluxetable}{lcccc} 
\tablewidth{0pt} 
\tablecaption{Reverberation Results From Kaspi et al.\ Dataset} 
\tablehead{ 
\colhead{Data} & 
\colhead{Spectral} &
\colhead{$\tau_{\rm cent} $} & 
\colhead{$\tau_{\rm peak} $}\\ 
\colhead{Subset} & 
\colhead{Feature} &
\colhead{(days) } &
\colhead{(days)} 
} 
\startdata
Entire Data Set & H$\alpha$ & 215.2 $^{+32.1}_{-27.8}$ &  217.4 $^{+20.0}_{-20.0}$ \\
Entire Data Set & H$\beta$  & 168.2 $^{+36.7}_{-21.4}$ &  177.4 $^{+60.0}_{-100.0}$ \\
Entire Data Set & H$\gamma$ & 197.6 $^{+37.9}_{-28.5}$ &  208.3 $^{+40.0}_{-80.0}$ \\
1993-1995 & H$\alpha$ & 199.1 $^{+4.8}_{-0.7}$ &  203.0 $^{+10.0}_{-10.0}$ \\
1993-1995 & H$\beta$  & 203.5 $^{+5.2}_{-17.4}$ &  202.9 $^{+0.0}_{-10.0}$ \\
1993-1995 & H$\gamma$ & 195.6 $^{+35.1}_{-27.6}$ &  219.5 $^{+21.0}_{-93.0}$ \\
1993 & H$\alpha$ & 15.6 $^{+0.6}_{-1.4}$ &  19.1 $^{+0.0}_{-3.0}$ \\
1993 & H$\beta$  & 12.2 $^{+1.5}_{-2.5}$ &  13.1 $^{+3.0}_{-2.0}$ \\
1994 & H$\alpha$ & 15.2 $^{+0.9}_{-1.6}$ &  16.0 $^{+0.0}_{-0.0}$ \\
1994 & H$\beta$  & 13.7 $^{+1.8}_{-0.8}$ &  16.0 $^{+0.0}_{-1.0}$ \\
1994 & H$\gamma$ & 10.0 $^{+5.3}_{-3.4}$ &  12.0 $^{+51.0}_{-9.0}$ \\
1995 & H$\alpha$ & 36.6 $^{+4.7}_{-3.3}$ &  44.0 $^{+0.0}_{-10.0}$ \\
1995 & H$\beta$  & 46.5 $^{+4.0}_{-2.9}$ &  44.0 $^{+8.0}_{-0.0}$ \\
1995 & H$\gamma$ & 67.4 $^{+7.7}_{-8.5}$ &  67.0 $^{+7.7}_{-8.5}$ 
\enddata
\label{Table:lags}
\end{deluxetable} 

\end{document}